\documentclass[%
 aip, pop, amsmath, amssymb, reprint, 
 amsmath,amssymb,
 reprint,%
]{revtex4-1}
\usepackage{xcolor} 
\usepackage{graphicx}
\usepackage{dcolumn}
\usepackage{bm}

\usepackage[utf8]{inputenc}
\usepackage[T1]{fontenc}
\usepackage{mathptmx}
\usepackage{etoolbox}

\makeatletter
\def\@email#1#2{%
 \endgroup
 \patchcmd{\titleblock@produce}
  {\frontmatter@RRAPformat}
  {\frontmatter@RRAPformat{\produce@RRAP{*#1\href{mailto:#2}{#2}}}\frontmatter@RRAPformat}
  {}{}
}%
\makeatother

\begin{document}

\preprint{}

\title{Generalized mixed variable-pullback scheme with non-ideal Ohm's law for electromagnetic gyrokinetic simulations}
\author{Zhixin Lu}
\affiliation{%
Max Planck Institute for Plasma Physics, Boltzmannstr. 2, Garching, 85748, Germany 
}%

\author{Guo Meng}
\affiliation{%
Max Planck Institute for Plasma Physics, Boltzmannstr. 2, Garching, 85748, Germany 
}%

\author{Roman Hatzky}
\affiliation{%
Max Planck Institute for Plasma Physics, Boltzmannstr. 2, Garching, 85748, Germany 
}%

\author{Eric Sonnendrücker}
\affiliation{%
Max Planck Institute for Plasma Physics, Boltzmannstr. 2, Garching, 85748, Germany 
}%

\author{Alexey Mishchenko}
\affiliation{%
Max Planck Institute for Plasma Physics, D-17491 Greifswald, Germany
}%

\author{Matthias Hoelzl}
\affiliation{%
Max Planck Institute for Plasma Physics, Boltzmannstr. 2, Garching, 85748, Germany 
}%

\date{\today}

\begin{abstract}
In this work, the non-ideal Ohm's law is integrated in the mixed variable-pullback scheme for the gyrokinetic particle simulations. This scheme captures the evolution of the symplectic solution of the gyrokinetic model accurately not only in the MHD limit but also in the electrostatic limit. This scheme also provides a pure symplectic ($v_\shortparallel $) scheme for electromagnetic gyrokinetic particle simulations without causing the traditional cancellation problem in the pure Hamiltonian scheme. Various  mixed variable schemes have been comprehensively analyzed for the 1D shear Alfv\`en wave problem with kinetic electrons, with the connection to the traditional pure Hamiltonian scheme and the symplectic scheme. It is demonstrated that the pure $v_\shortparallel $ form with the non-ideal Ohm's law has comparable performance to the widely used mixed variable-pullback scheme with the ideal Ohm's law. The mixed variable-pullback scheme without Ohm's law is also proposed as a feasible improvement of the traditional pure Hamiltonian scheme with minimum modification and considerable performance improvement in terms of a marker number reduction. 
\end{abstract}

\maketitle

\section{Introduction}
The electromagnetic gyrokinetic particle simulation has played a crucial role in identifying the fundamental physics in toroidally confined plasmas, especially for high-$\beta$ burning plasmas \cite{mishchenko2023numerical}. 
To enhance the capability of the electromagnetic gyrokinetic simulations, various physics models and numerical schemes have been developed, such as the the iterative $p_\shortparallel $ scheme \cite{chen2007electromagnetic}, the mixed variable/pullback scheme \cite{mishchenko2019pullback}, the implicit scheme \cite{lu2021development,sturdevant2021verification}, the GK-E\&B model and its implementation \cite{chen2021gyrokinetic,rosen2022and}, the conservative scheme \cite{bao2018conservative} and the noise reduction scheme \cite{hatzky2019reduction}. The Hamiltonian and symplectic schemes have also been studied in gyrokinetic continuum codes GENE \cite{gorler2016intercode} and GKeyll \cite{hakim2020continuum}. In addition, advanced mathematical models have been developed, such as the geometric particle-in-cell method in the GEMPIC code \cite{kraus2017gempic,meng2025geometric}. These models are key ingredients in modern gyrokinetic codes for the  prediction and interpretation of experimental results  \cite{lanti2020orb5,kleiber2024euterpe,taimourzadeh2019verification}.  

The mixed variable–pullback scheme is regarded as one of the most efficient approaches for electromagnetic simulations \cite{mishchenko2014pullback} and has been implemented in various codes such as ORB5 \cite{mishchenko2019pullback}, EUTERPE \cite{kleiber2024euterpe}, TRIMEG \cite{lu2023full,lu2024gyrokinetic}, XGC \cite{cole2021tokamak,hager2022electromagnetic}, and GTS \cite{startsev2024verification}. This scheme requires specifying an equation for the symplectic part of the perturbed vector potential $\delta A_\shortparallel ^{\rm s}$. Traditionally, the ideal Ohm’s law has been used, which offers particular efficiency in the MHD limit but provides little advantage in the electrostatic limit. While the symplectic $v_\shortparallel $ scheme with Ohm's law has also been discussed and/or implemented in the  gyrokinetic continuum codes \cite{hakim2020continuum,gorler2016intercode}, the comparison of various schemes has not been reported, and it is unclear whether the pure $v_\shortparallel $ form is as good as or even better than the mixed variable scheme. In this work, the impact of the pullback operation on the time evolution of the symplectic and Hamiltonian part of $\delta A_\shortparallel $ is analyzed for the first time in detail for various $\beta$ values. Specifically, this work addresses the generalized mixed variable-pullback scheme with non-ideal Ohm's law and compares it with other schemes. The remainder of this article is organized as follows. Section \ref{sec:model} presents the models and equations of the traditional mixed variable-pullback scheme and the generalized one using the non-ideal Ohm's law. Section \ref{sec:theory} describes the theoretical analyses. In Section \ref{sec:results}, simulation results are shown for evaluating the performance of different schemes. Finally, conclusions are made in Section~\ref{sec:summary}.

\section{Physics model and equations}
\label{sec:model}
\subsection{Traditional mixed variable-pullback scheme with ideal Ohm's law}
Using the mixed variable scheme, the parallel component of the perturbed magnetic potential $\delta A_\shortparallel $ is decomposed into a symplectic part and a Hamiltonian part \cite{mishchenko2014pullback},
\begin{equation}
\label{eq:AsAh}
    \delta A_\shortparallel  =\delta A_\shortparallel ^{\rm{s}} + \delta A_\shortparallel ^{\rm{h}} \;\;, 
\end{equation}
where $\delta A_\shortparallel ^{\rm{s}}$ and $\delta A_\shortparallel ^{\rm{h}}$ are the symplectic part and the Hamiltonian part, respectively. 
The shifted parallel velocity coordinate of the gyrocenter $u_\shortparallel $ is defined as 
\begin{equation}
    u_\shortparallel =v_\shortparallel +\frac{q_s}{m_s}\langle\delta A_\shortparallel ^{\rm{h}}\rangle\;\;,
\end{equation}
where $v_\shortparallel $ is the parallel velocity, $q_s$ and $m_s$ are the charge and mass of species $s$, respectively, the subscript ``$s$'' represents the different particle species, and $\langle\ldots\rangle$ indicates the gyro average. In the pure $p_\shortparallel $ form, $\delta A_\shortparallel ^{\rm s}=0$ and $p_\shortparallel =v_\shortparallel +(q_s/m_s)\delta A_\shortparallel $ is introduced. In this work, we adopt the general definition of $u_\shortparallel $ that can be reduced to that of $p_\shortparallel $ as $\delta A_\shortparallel ^{\rm s}=0$.

The gyrocenter equations of motion are consistent with previous work \cite{mishchenko2019pullback,hatzky2019reduction,lanti2020orb5,mishchenko2023global,kleiber2024euterpe},
\begin{eqnarray}
\label{eq:gcmotion1}
\dot{\bm R}_0 
  &=& u_\shortparallel  {\boldsymbol b}^*_0 + \frac{m\mu}{qB^*_\shortparallel } {\boldsymbol b}\times\nabla B \;\;, \\
\label{eq:gcmotion2}
  \dot u_{\shortparallel ,0}
  &=& -\mu {\boldsymbol b}^*_0\cdot \nabla B \;\;,
\end{eqnarray}
\begin{eqnarray}
\label{eq:gcmotion3}
  \delta\dot{\boldsymbol R}
  &=& \frac{{\boldsymbol b}}{B^*_\shortparallel }\times \nabla \langle \delta\Phi -u_\shortparallel  \delta A_\shortparallel \rangle 
  -\frac{q_s}{m_s}\langle\delta A^{\rm h}_\shortparallel \rangle {\boldsymbol b}^*\;\;, 
  \\
\label{eq:gcmotion4}
  \delta \dot u_\shortparallel 
  &=&  -\frac{q_s}{m_s} \left({\boldsymbol b}^*\cdot\nabla\langle\delta\Phi-u_\shortparallel \delta A^{\rm{h}}_\shortparallel \rangle +\partial_t\langle\delta A_\shortparallel ^{\rm{s}}\rangle \right) \nonumber\\
  &&-\frac{\mu}{B^*_\shortparallel }{\boldsymbol b}\times\nabla B\cdot\nabla\langle\delta A_\shortparallel ^{\rm{s}}\rangle \;\;,  
\end{eqnarray}
where $\delta\Phi$ is the perturbed electrical potential, the magnetic moment $\mu=v_\perp^2/(2B)$, ${\boldsymbol b}^*={\boldsymbol b}_0^*+\nabla\langle\delta A_\shortparallel ^{\rm s}\rangle\times{\boldsymbol b}/B_\shortparallel ^*$, $\langle\ldots\rangle$ denotes the gyro average, ${\boldsymbol b}^*_0={\boldsymbol b}+(m_s/q_s)u_\shortparallel \nabla\times{\boldsymbol b}/B_\shortparallel ^*$, ${\boldsymbol b}={\bm B}/B$, $B_\shortparallel ^*=B+(m_s/q_s)u_\shortparallel {\boldsymbol b}\cdot(\nabla\times{\boldsymbol b})$, $\boldsymbol{B}$ is the equilibrium magnetic field. 

The perturbed distribution function is solved by following the marker trajectory,
\begin{eqnarray}
\label{eq:dfdt}
    \frac{{\rm d}\delta f}{{\rm d}t} = -\delta\dot{\bm R} \cdot\nabla f_0 -\dot v_\shortparallel \frac{\partial }{\partial v_\shortparallel }f_0 \;\;,
\end{eqnarray}
where the equilibrium distribution $f_0$ is the steady state solution that satisfies ${\rm d}f_0/{\rm d}t|_0=0$ along the unperturbed trajectory, and it is typically assumed to be the Maxwellian distribution function when the neoclassical physics is not considered in toroidally confined plasmas. 

The linearized quasi-neutrality equation with the long-wavelength approximation is as follows, 
\begin{equation}
\label{eq:poisson0}
    -\nabla\cdot\left( \sum_s\frac{q_s n_{0s}}{B\omega_{{\rm c}s}} \nabla_\perp\delta\Phi \right) = \sum_s q_s \delta n_{s,v} \;\;,
\end{equation}
where the gyrocenter density $\delta n_{s,v}$ is calculated using $\delta f_s({\boldsymbol R},v_\shortparallel ,\mu)$ (denoted as $\delta f_{s,v}$), namely, $\delta n_{s,v}({\boldsymbol{x} })=\int {\rm d}^6 z\,\delta f_{s,v}\delta(\boldsymbol{R}  + \boldsymbol{\rho} - \boldsymbol{x} )$. Here, $\boldsymbol{x} $ and $\boldsymbol{R} $ denote the particle position vector and gyrocenter position vector, respectively, and $\boldsymbol{\rho}$ represents the Larmor radius vector.
In Eq.~\eqref{eq:poisson0}, $\omega_{{\rm c}s}$ is the cyclotron frequency of species $s$. 
$\delta f_{s,v}$ is obtained from $\delta f_{s,u}$ with the linear approximation of the pullback scheme,
\begin{eqnarray}
    &&\delta f_{s,v} = \delta f_{s,u} +  \frac{q_s\left\langle\delta A^{\rm{h}}_{\shortparallel } \right\rangle}{m_s}\frac{\partial f_{0s}}{\partial v_\shortparallel } \nonumber \\
    &&\xrightarrow[f_{0s}=f_{\rm{M}}]{\text{Maxwellian}}
     \delta f_{s,u} -  \frac{ v_\shortparallel }{T_s}  q_s\left\langle\delta A^{\rm{h}}_{\shortparallel } \right\rangle f_{0s} \;\;
\end{eqnarray}
produced from the general form of the nonlinear pullback scheme \cite{hatzky2019reduction},
\begin{eqnarray}
    & f_{s,v} (v_\shortparallel ) = f_{s,u} \left(v_\shortparallel +\frac{q_s}{m_s}\langle\delta A_\shortparallel ^{\rm h}\rangle\right)\;\;.
\end{eqnarray}

Amp\`ere's law is expressed as 
\begin{eqnarray}
\label{eq:ampere_mv_deltaf_exact}
    -\nabla^2_\perp\delta A_{\shortparallel }^{\rm{h}}
    +\sum_s\mu_0\frac{q_s^2}{T_s}\int \mathrm{d}^6z\, v_\shortparallel ^2 f_{0s} \langle \delta A_{\shortparallel }^{\rm{h}} \rangle \delta(\boldsymbol{R}  + \boldsymbol{\rho} - \boldsymbol{x} ) \nonumber\\
    =\nabla^2_\perp\delta A_{\shortparallel }^{\rm{s}} 
    + \mu_0\sum_s q_s  \int \mathrm{d}^6z \, v_\shortparallel  \delta f_{s,u}(u_\shortparallel )\delta(\boldsymbol{R}  + \boldsymbol{\rho} - \boldsymbol{x} )   \;\;.
\end{eqnarray}

In the previous work, the ideal Ohm's law is adopted to determine $\delta A_\shortparallel ^{\rm{s}}$, 
\begin{equation}
    \partial_t\delta A_\shortparallel ^{\rm{s}}+\partial_\shortparallel \delta\Phi = 0\;\;,
\label{eq:ohm_law0}
\end{equation}
where the parallel derivative is defined as $\partial_\shortparallel =\boldsymbol{b}\cdot\nabla$, $\boldsymbol{b}=\boldsymbol{B}/B$ as already defined following Eq.~\eqref{eq:gcmotion4}. 
It is believed that Eq.~\eqref{eq:ohm_law0} makes the gyrokinetic simulation efficient in the MHD limit. 
However, in the electrostatic limit  ($\delta A\approx0$), Eq.~\eqref{eq:ohm_law0} leads to a large amplitude of $\delta A_\shortparallel ^{\rm{s}}$ , which in turn requires a correspondingly large $\delta A_\shortparallel ^{\rm{h}}$ to compensate and recover the correct physics. As a result, the mixed variable  scheme loses its original advantages in reducing $\delta A^{\rm h}$ and the simulation noise unless a ``pullback'' operation is introduced as follows. 

At the end of each step, the solution is pulled back to the symplectic form by converting  $(\delta f_{s,u},u_\shortparallel )$ to $(\delta f_{s,v},v_\shortparallel )$, 
\begin{eqnarray}
\label{eq:pullback_f}
    \delta f^{\rm new}_{s,u} &=& \delta f^{\rm old}_{s,u} +  \frac{q_s\left\langle\delta A^{\rm{h,old}}_{\shortparallel } \right\rangle}{m_s}\frac{\partial f_{0s}}{\partial v_\shortparallel } \nonumber \\
&&    \xrightarrow[f_{0s}=f^{\rm old}_{\rm{M}}]{\text{Maxwellian}} 
     \delta f_{s,u}^{\rm old} -  \frac{ v_\shortparallel }{T_s}  q_s\left\langle\delta A^{\rm{h,old}}_{\shortparallel } \right\rangle f_{0s} \;\;,\\
\label{eq:pullback_v}
     u^{\rm new}_{\shortparallel } &=& u^{\rm old}_{\shortparallel }  -\frac{q_s}{m_s} \langle\delta A_\shortparallel ^{\rm h,old}\rangle \;\;, \\
\label{eq:pullback_A}
     \delta A_\shortparallel ^{\rm s,new} &=& \delta A_\shortparallel ^{\rm s,old} + \delta A_\shortparallel ^{\rm h,old}\;\;,\;\;
     \delta A_\shortparallel ^{\rm h,new} = 0 \;\;.
\end{eqnarray}
In this work, we employ a 4th-order Runge-Kutta time integrator, and the pullback operation is applied at the end of each full time step, which consists of four sub-stages. 

\subsection{Non-ideal Ohm's law}
In this work, we use the non-ideal Ohm's law to determine $\partial_t\delta A_\shortparallel ^{\rm{s}}$. We first derive the rigorous non-ideal Ohm's law. Then, different approximations are made to recover the ideal Ohm's law and other schemes. The derivation of the non-ideal Ohm's law is similar to the $E_\shortparallel $ equation in the GK-E\&B model reported previously \cite{chen2021gyrokinetic,rosen2022and}. However, we adopt $\delta A_\shortparallel $ and $\delta\Phi$ instead of the electric field $E$ and the magnetic field $B$ to derive the $\partial _t\delta A$ equation using the $\delta f$ method. 

For the sake of simplicity, we derive the non-ideal Ohm's law in pure $v_\shortparallel $ form corresponding to $\delta A_\shortparallel ^{\rm h}=0$,  $\delta A_\shortparallel =\delta A_\shortparallel ^{\rm s}$. Then Eqs.~\eqref{eq:gcmotion1}--\eqref{eq:gcmotion4} are reduced to 
\begin{eqnarray}
 \dot{\boldsymbol R}_0 
  &=& v_\shortparallel  {\boldsymbol b} +{\bm v}_d  \\
  {\bm v}_d 
  &=&
  \frac{m_sv_\shortparallel ^2}{q_sB_\shortparallel ^*}\nabla\times{\boldsymbol b}+\frac{m\mu}{qB^*_\shortparallel } {\boldsymbol b}\times\nabla B \;\;, 
  \\
  \dot v_{\shortparallel ,0}
  &=& -\mu {\boldsymbol b}^*_0\cdot \nabla B \;\;,
  \\
  \delta\dot{\boldsymbol R}
  &=& \frac{{\boldsymbol b}}{B^*_\shortparallel }\times \nabla \langle \delta\Phi -v_\shortparallel  \delta A_\shortparallel \rangle \;\;, 
  \\
  \delta \dot v_\shortparallel 
  &=&  -\frac{q_s}{m_s} \left({\boldsymbol b}^*\cdot\nabla\langle\delta\Phi\rangle +\partial_t\langle\delta A_\shortparallel  \rangle \right)  \nonumber\\
  &&-\frac{\mu}{B^*_\shortparallel }{\boldsymbol b}\times\nabla B\cdot\nabla\langle\delta A_\shortparallel \rangle \;\;.  
\end{eqnarray}
Amp\`ere's law in $v_\shortparallel $ space is given by 
\begin{equation}
    -\nabla^2_\perp\delta A_\shortparallel  = \mu_0 \delta j_{\shortparallel ,v} \;\;,
\end{equation}
where $\delta j_{\shortparallel ,v}({\boldsymbol{x} })=\sum_s q_s \int {\rm d}^6 z\, \delta f_{s,v}\delta(\boldsymbol{R}  + \boldsymbol{\rho} - \boldsymbol{x} )v_\shortparallel $.
The time derivative yields
\begin{equation}
\label{eq:dAdt0}
    -\nabla^2_\perp\partial_t\delta  A_\shortparallel  = \mu_0  \partial_t \delta j_{\shortparallel ,v} \;\;. 
\end{equation}
By  taking the $qv_\shortparallel $ moment of the nonlinear gyrokinetic equation
\begin{eqnarray}
    \partial_t\delta f + {\bm v}\cdot\nabla \delta f +\dot v_\shortparallel \frac{\partial }{\partial v_\shortparallel }\delta f =
    -\delta {\bm v}\cdot\nabla f_0 - \delta \dot v_\shortparallel  \frac{\partial }{\partial v_\shortparallel } f_0 \;\;,
\end{eqnarray}
we obtain
\begin{eqnarray}
\label{eq:djdt0}
\partial_t\delta j_\shortparallel  =\sum_s q_s\left[
    \left\langle J_0 \delta f\frac{\partial}{\partial v_\shortparallel }(\dot v_\shortparallel  v_\shortparallel ) \right\rangle_v
    -\langle J_0 {\bm v}\cdot \nabla(v_\shortparallel \delta f) \rangle_v \right. \nonumber\\
    \left.
    -\langle J_0 v_\shortparallel \delta{\bm v}\cdot\nabla f_0 \rangle_v
    -\left\langle J_0 v_\shortparallel \delta \dot v_\shortparallel \frac{\partial f_0}{\partial v_\shortparallel } \right\rangle_v
\right]\;\;,
\end{eqnarray}
where $\langle A\rangle_v=\int {\rm d}v^3 A$, the Bessel function $J_0$ takes into account the finite Larmor radius effect, and the identity $\nabla v_\shortparallel =0$ and the phase-space incompressibility $\partial_tB_\shortparallel ^*+\nabla\cdot(B_\shortparallel ^*{\bm \dot R})+\partial(B_\shortparallel ^*\dot v_\shortparallel )/\partial v_\shortparallel =0$ (see, e.g., [\!\!\citenum{meng2025geometric}]) have been used. 
Using the gyrocenter equations of motion, Eq.~\eqref{eq:djdt0} yields
\begin{eqnarray}
\label{eq:djdt1}
\partial_t\delta j_\shortparallel  &=& -\sum_s \frac{q_s}{m_s}\left[
    \partial_\shortparallel  \delta P_\shortparallel 
    +\frac{\nabla B}{B}\cdot(\delta\tilde{\bf P}_\perp-\delta P_\shortparallel {\bf b}) \right. \nonumber \\
& & + m_s\nabla\cdot \langle J_0 v_\shortparallel ({\bm v}_d + \delta {\bm v})\delta f \rangle_v   
-q_s\langle J_0 \delta f E_{\shortparallel 0}\rangle_v  
\nonumber\\
    &&\left.
    +m_s\langle J_0 v_\shortparallel \delta {\bm v}\cdot\nabla f_0 \rangle_v
    -q_s \langle J_0f \delta E_\shortparallel  \rangle_v
\right]\;\;,
\end{eqnarray}
where $\delta P_\shortparallel =\langle J_0 \delta f mv_\shortparallel ^2\rangle_v $, $\delta\tilde{{\bf P}}_\perp=\langle J_0 \delta f m\mu B {\bf b}_0^*\rangle_v$, $\delta E_\shortparallel =-\partial_t\delta A_\shortparallel -\partial_\shortparallel \delta \Phi$, $E_\shortparallel =E_{0\shortparallel }+\delta E_\shortparallel $, $E_0$ is the equilibrium electric field, and we have made use of $\partial_\shortparallel  B/B+\nabla\cdot{\bm b}=0$.  
The non-ideal Ohm's law is obtained from Eqs.~\eqref{eq:dAdt0} and \eqref{eq:djdt1},
\begin{eqnarray}
\label{eq:dAdt_complete}
    \bigg(
    \nabla_\perp^2  &-& \sum_s M_{ds} 
    \bigg) \partial_t\delta A_\shortparallel  =
     \sum_sM_{ds} \partial_\shortparallel \delta\Phi \nonumber\\
     &&+\sum_s\frac{1}{d_{s0}^2 n_{s0} q_s} 
    \left[
    \partial_\shortparallel  \delta P_\shortparallel 
    +(\delta P_\shortparallel -\delta P_\perp) \nabla\cdot{\bm b}\right. \nonumber\\
    &&
    +m_s\nabla\cdot \langle J_0 v_\shortparallel ({\bm v}_d + \delta {\bm v})\delta f \rangle_v \nonumber \\
    && \left. -q_s\langle J_0 \delta f E_{\shortparallel 0}\rangle_v 
    +m_s\langle J_0 v_\shortparallel \delta {\bm v}\cdot\nabla f_0 \rangle_v
    \right] \;\;, \\
    M_{ds} Y & = &
    \frac{\mu_0 e_s^2}{m_s}\langle J_0 f Y \rangle_v 
    \approx \frac{1}{d_s^2} Y \;, \; Y=\{\partial_t\delta A_\shortparallel , \delta\Phi \} \;,
\label{eq:Mds}
\end{eqnarray}
where the skin depth of species ``s'' is defined as $d_s=c/\omega_{ps}$, $\omega_{ps}=\sqrt{n_sq_s^2/(m_s\varepsilon_0)}$,  $d_{s0}=c/\omega_{ps0}$, $\omega_{ps0}=\sqrt{n_{0s}q_s^2/(m_s\varepsilon_0)}$, $n_s=\langle f\rangle_v$, $n_{0s}=\langle f_0\rangle_v$. 
The nonlinear electron skin depth term originates from the velocity integral in Eq.~\eqref{eq:Mds}. In the 1D problem considered in this work, it is calculated at each step. For 2D and 3D problems solved with the finite element method, directly calculating $M_{ds}Y$ requires generating the mass matrix at every step and can be expensive for large-scale problems. A common approach is therefore to first compute the lowest-order solution and then use the iterative scheme \cite{chen2007electromagnetic,hatzky2019reduction,lu2023full} to get the accurate solution.
To get a good approximation of Eq.~\eqref{eq:dAdt_complete}, by separating the nonlinear skin depth term into an equilibrium and perturbed parts, ignoring the gyroaverage in the $M_{ds}$ term, and keeping the dominant terms in ${\bf b}_0^*\cdot B$, we obtain
\begin{eqnarray}
\label{eq:dAdt_complete_dominant}
    &&\left(
    \nabla_\perp^2-\sum_s\frac{1}{d_{s0}^2} 
    \right) \partial_t\delta A_\shortparallel  =
     \sum_s\frac{1}{d_{s0}^2} \partial_\shortparallel \delta\Phi \nonumber\\
     &&+\sum_s\frac{1}{d_s^2 n_s q_s} 
    \left[
    \partial_\shortparallel  \delta P_\shortparallel 
    +(\delta P_\shortparallel -\delta P_\perp) \nabla\cdot{\bm b}\right. \nonumber\\
    &&
    +m_s\nabla\cdot \langle J_0 v_\shortparallel ({\bm v}_d + \delta {\bm v})\delta f \rangle_v \nonumber \\
    && \left. -q_s\langle J_0 \delta f E_\shortparallel \rangle_v 
    +m_s\langle J_0 v_\shortparallel \delta {\bm v}\cdot\nabla f_0 \rangle_v
    \right] \;\;,
\end{eqnarray}
where $\delta P_\perp=\langle J_0 \delta f m\mu B \rangle_v $.
Equation~\eqref{eq:dAdt_complete_dominant} is closely linked to the $E_\shortparallel $ equation in the GK-E\&B model \cite{chen2021gyrokinetic} that is derived from the full $f$ model. From the $E_\shortparallel $ equation \cite{chen2021gyrokinetic}, by separating the distribution $f=f_0+\delta f$ and the equilibrium equation, representing $E$ and $B$ using $\delta A_\shortparallel $ and $\delta\Phi$, and with the simplification $\nabla\times\delta B\approx-\nabla_\perp^2\delta A_\shortparallel $, Eq.~\eqref{eq:dAdt_complete_dominant} can be also obtained.

The cancellation problem potentially exists in Eq.~\eqref{eq:dAdt_complete} if it is not treated properly. In the $\delta E_\shortparallel \rightarrow0$ limit, the skin depth terms $\partial_t\delta{A}_\shortparallel \sum_s (1/d_s^2)$ and $\partial_\shortparallel \delta\Phi \sum_s (1/d_s^2)$ need to cancel each other. It can be achieved by calculating these two terms accurately directly, or using an iterative method \cite{hatzky2019reduction,mishchenko2019pullback,lu2023full}. In addition, these two terms can be treated both with the analytical expression of $\sum_s (1/d_s^2)$ to achieve the complete cancellation, while in the traditional mixed variable-pullback scheme Eq. \eqref{eq:ampere_mv_deltaf_exact} or the pure $p_\shortparallel $ scheme, the current in $u_\shortparallel $ space can only be calculated using markers. 

\subsection{Mixed variable-pullback scheme with non-ideal Ohm's law for $\delta A_\shortparallel ^{\rm s}$}
In principle, Eq.~\eqref{eq:dAdt_complete} can be used as the equation for $\delta A_\shortparallel ^{\rm s}$,  
\begin{eqnarray}
\label{eq:dAsdt_complete}
    \left(
    \nabla_\perp^2-\sum_s\frac{1}{d_s^2} 
    \right) \partial_t\delta A_\shortparallel ^{\rm s} &=& {\rm R.H.S} \;\;,
\end{eqnarray}
where ${\rm R.H.S}$ denotes the right-hand-side terms in Eq.~\eqref{eq:dAdt_complete} where $\delta f=\delta f_v$ that can be pulled back from $\delta f_u$.  
While Eq.~\eqref{eq:dAsdt_complete} is complete and rigorous, numerical errors in the time integrator and the field solver exist. Consequently, as $\delta A_\shortparallel ^{\rm h}$ is solved from $\delta A_\shortparallel ^{\rm s}$, we have $\delta A_\shortparallel ^{\rm h}=\varepsilon$, where $\varepsilon$ is from numerical errors.

The more general way is to use the simplified version of Eq.~\eqref{eq:dAdt_complete} for $\delta A_\shortparallel ^{\rm s}$. 
Equation~\eqref{eq:dAdt_complete} can be simplified by keeping the dominant terms and be used as an equation for $\delta A_\shortparallel ^{\rm s}$. In the slab geometry with a uniform magnetic field and zero equilibrium electric field ${\bm E}_0=0$, Eq.~\eqref{eq:dAdt_complete} is reduced to
\begin{eqnarray}
\label{eq:dAdt_slab}
    \left(
    \nabla_\perp^2-\sum_s\frac{1}{d_s^2} 
    \right) \partial_t\delta A_\shortparallel ^{\rm s} &=& 
     \sum_s\frac{1}{d_s^2} \partial_\shortparallel \delta\Phi+\sum_s\frac{1}{d_s^2 n_s q_s} 
    \partial_\shortparallel  \delta P_\shortparallel \;\;, \nonumber \\
\end{eqnarray}
which describes the balance between the inertia (the $\nabla^2_\perp\delta A$ term), the parallel acceleration, and the parallel pressure gradient.

Another choice is to keep the inertia and the parallel acceleration,
\begin{eqnarray}
\label{eq:dAdt_slab_nopressure}
    \left(
    \nabla_\perp^2-\sum_s\frac{1}{d_s^2} 
    \right) \partial_t\delta A_\shortparallel ^{\rm s} &=& 
     \sum_s\frac{1}{d_s^2} \partial_\shortparallel \delta\Phi\;\;.
\end{eqnarray}
Although Eq.~\eqref{eq:dAdt_slab_nopressure} is significantly simplified, it recovers the MHD limit in the small electron skin depth limit ($|k_\perp d_{\rm e}|\ll1)$ in Eq.~\eqref{eq:ohm_law0}. In addition, in the large electron skin depth limit ($|k_\perp d_{\rm e}|\ll1$), 
\begin{eqnarray}
\label{eq:dAdt_slab_largede}
    |\partial_t\delta A_\shortparallel ^{\rm s}| 
    &\approx& 
     \left|\frac{1}{-k_\perp^2-\sum_s\frac{1}{d_s^2} }\sum_s\frac{1}{d_s^2} \partial_\shortparallel \delta\Phi\right| 
     \ll
     |\delta\Phi|
     \;\;,
\end{eqnarray}
which already mitigates the artificial large amplitude of $\delta A_\shortparallel ^{\rm s}$  in the traditional mixed variable-pullback scheme in the large $|k_\perp d_{\rm e}|$ or the electrostatic limit.   

As a further simplification, since electrons mainly contribute to the current, the contribution from other species can be ignored or approximated using the treatment for electrons (zero Larmor radius effect). 

The mixed variable-pullback scheme with non-ideal Ohm's law consists of the following steps.
\begin{enumerate}
    \item Solve the non-ideal Ohm's law, either the complete or the simplified forms in Eqs.~\eqref{eq:dAsdt_complete}, \eqref{eq:dAdt_slab}, or \eqref{eq:dAdt_slab_nopressure}.
    \item Solve the Quasi-neutrality equation \eqref{eq:poisson0} and Amp\`ere's law in Eq.~\eqref{eq:ampere_mv_deltaf_exact}
    \item Solve the gyrocenter equations of motion in Eqs.~\eqref{eq:gcmotion1}--\eqref{eq:gcmotion4} and the $\delta f$ equation in Eq.~\eqref{eq:dfdt}. 
    \item Pull back the parallel velocity, the perturbed distribution function and the perturbed field $(\delta A_\shortparallel ^{\rm s},\delta A_\shortparallel ^{\rm h})$ in Eqs.~\eqref{eq:pullback_f}--\eqref{eq:pullback_A}. 
\end{enumerate}

Five schemes are listed in Tab.~\ref{tab:various_schemes}. All the schemes can be described by the above steps but with a reduction of $\delta A_\shortparallel ^{\rm h}=0$ in the pure $v_\shortparallel $ scheme or $\delta A_\shortparallel ^{\rm s}=0$ in the pure $p_\shortparallel $ scheme. In both the pure $v_\shortparallel $ scheme and the $p_\shortparallel $ scheme, the pullback operation is not employed. 

\begin{table*}
\centering
\begin{tabular}{ c c|c|c|c  } 
 \hline
   & Method    &  GC Eqs. & Field Eqs. & Pullback\\
\hline
 1 & MVPB, ideal Ohm & \eqref{eq:gcmotion1}--\eqref{eq:gcmotion4} with $\delta A_\shortparallel ^{\rm s}$, $\delta A_\shortparallel ^{\rm h}$ & \eqref{eq:poisson0}, \eqref{eq:ampere_mv_deltaf_exact}, \eqref{eq:ohm_law0} & yes \\ 
 \hline
 2a & MVPB, non-ideal Ohm & \eqref{eq:gcmotion1}--\eqref{eq:gcmotion4} with $\delta A_\shortparallel ^{\rm s}$, $\delta A_\shortparallel ^{\rm h}$ &\eqref{eq:poisson0}, \eqref{eq:ampere_mv_deltaf_exact}, \eqref{eq:dAsdt_complete} & yes \\
 \hline
 2b & Pure $v_\shortparallel $, non-ideal Ohm & \eqref{eq:gcmotion1}--\eqref{eq:gcmotion4} with $\delta A_\shortparallel ^{\rm h}=0$  & \eqref{eq:poisson0}, \eqref{eq:ampere_mv_deltaf_exact} with $\delta A_\shortparallel ^{\rm h}=0$, \eqref{eq:dAdt_complete} & no \\ 
 \hline
 3a & MVPB, no Ohm & \eqref{eq:gcmotion1}--\eqref{eq:gcmotion4} with $\delta A_\shortparallel ^{\rm s}$, $\delta A_\shortparallel ^{\rm h}$  & \eqref{eq:poisson0}, \eqref{eq:ampere_mv_deltaf_exact} & yes\\
 \hline
 3b & Pure $p_\shortparallel $, no Ohm & \eqref{eq:gcmotion1}--\eqref{eq:gcmotion4} with $\delta A_\shortparallel ^{\rm s}=0$ & \eqref{eq:poisson0}, \eqref{eq:ampere_mv_deltaf_exact} with $\delta A_\shortparallel ^{\rm s}=0$ & no \\ 
 \hline
\end{tabular}
 \caption{The mixed variable-pullback scheme (MVPB) and the reduction to the pure $v_\shortparallel $ and pure $p_\shortparallel $ schemes. }
 \label{tab:various_schemes}
\end{table*}

\section{Theoretical analyses}
\label{sec:theory}
The uniform plasma in a slab geometry with a uniform equilibrium magnetic field \cite{kleiber2024euterpe,bao2018conservative} is considered for the theoretical analyses. We consider drift kinetic electrons as the only kinetic species. The dominant ion response is the polarization density perturbation in the quasi-neutrality equation. The normalized quasi-neutrality equation,  the Amp\`ere's law, and the $\delta f$ equation are
\begin{eqnarray}
\label{eq:qn_slab}
    \nabla_\perp^2\delta\bar\Phi & =& \frac{1}{\bar\rho_N^2}\frac{\langle \delta f_{\rm e}\rangle_v}{n_0} \;\;, \\
\label{eq:ampere_slab}
    \nabla_\perp^2\delta\bar A_\shortparallel  &=& \frac{\beta}{\bar\rho_N^2}\frac{\langle v_\shortparallel \delta f_{\rm e}\rangle_v}{n_0} \;\;, \\
\label{eq:dfdt_slab}
    \frac{\rm d}{{\rm d}t}\delta f &=&\partial_t \delta f+v_\shortparallel  \partial_\shortparallel \delta f+\dot v_\shortparallel \partial_{v_\shortparallel }\delta f \nonumber \\
    &=& -\delta v_\shortparallel \frac{\partial f_0}{\partial x} -\delta\dot{v}_\shortparallel \frac{\partial}{\partial v_\shortparallel }f_0\;\;,
\end{eqnarray}
where $\langle\rangle_v$ denotes integration over velocity space, $\beta=v_N^2/v_{\rm A}^2$, $v_{\rm A}=B/\sqrt{\mu_0 m_N n_N}$ is the Alfv\'en velocity, $\rho_N=m_Nv_N/(eB_{\rm ref})$, $v_N=\sqrt{2T_N/m_N}$, the reference magnetic field $B_{\rm ref}=1$, $m_N$ is the ion mass, $T_N$ is the plasma temperature, we have assumed that the temperature is the same for electrons and ions and $x$ is the parallel coordinate. Given $\beta$ and $\rho_N$, the plasma temperature and the density can be determined, and vice versa. The subscript ``$N$'' denotes the normalization unit and the variables $(\rho_N, v_\shortparallel , t, \delta\Phi, \delta{A}_\shortparallel )$ are normalized to $(R_N=1 {\rm m}, v_N, R_N/v_N, m_Nv_N^2/e, m_N v_N/e)$.
To derive the linear dispersion relation, the linear terms in Eq.~\eqref{eq:dfdt_slab} are kept, 
\begin{eqnarray}
\label{eq:deltaf_slab}
    \delta f&=&-\frac{q_{\rm e}v_\shortparallel}{T}\frac{\omega\delta A_\shortparallel -k_\shortparallel \delta\Phi}{\omega-k_\shortparallel v_\shortparallel } f_0 \;\;.
\end{eqnarray}
We readily obtain 
$\delta n/n_0\equiv\langle\delta f \rangle/n_0=-2\bar{e}_{\rm e}/(\bar{T}_{\rm e}k_\shortparallel )[1+\alpha Z(\alpha)](k_\shortparallel \delta\Phi-\omega\delta A_\shortparallel )$, 
$\delta (nv_\shortparallel )/n_0\equiv\langle\delta f v_\shortparallel \rangle/n_0=-2\bar{e}_{\rm e}\alpha/(\sqrt{\bar{T}_{\rm e}\bar{m}_{\rm e}}k_\shortparallel )[1+\alpha Z(\alpha)](k_\shortparallel \delta\Phi-\omega\delta A_\shortparallel )$ and the dispersion relation 
\begin{eqnarray}
\label{eq:dispersion_slab0}
    D= 1-\frac{2\beta}{k_\perp^2\rho_N^2 \bar{T}} [1+\alpha Z(\alpha)] \left(
    \frac{\omega^2}{k_\shortparallel ^2} -\frac{1}{\beta}
    \right)\;\;,
\end{eqnarray}
where $\alpha=(\omega/k_\shortparallel )\sqrt{\bar{m}_{\rm e}/\bar{T}_{\rm e}}$. 

Equation~\eqref{eq:dAdt_slab} can be used to replace Eq.~\eqref{eq:ampere_slab} by replacing $\delta A_\shortparallel ^{\rm s}$ with $\delta A_\shortparallel $, to yield the same dispersion relation Eq.~\eqref{eq:dispersion_slab0}, 
\begin{eqnarray}
\label{eq:dAdt_slab1d}
    \left(
    \nabla_\perp^2-\frac{\beta\bar{q}_{\rm e}^2}{\rho_N^2 \bar{m}_{\rm e}} 
    \right) \partial_t\delta A_\shortparallel  &=& 
     \frac{\beta\bar{q}_{\rm e}^2}{\rho_N^2 \bar{m}_{\rm e}}  \partial_\shortparallel \delta\Phi
     +\frac{\beta\bar{q}_{\rm e}}{\rho_N^2 \bar{m}_{\rm e} n_0} 
    \partial_\shortparallel  \langle \bar{v}_\shortparallel ^2\rangle_v\;\;, \nonumber \\
\end{eqnarray}
where $\langle \bar{v}_\shortparallel ^2\rangle_v/n_0=-[\bar{q}_{\rm e}/(\bar{m_{\rm e}}\bar{k}_\shortparallel )][1+\alpha^2(1+\alpha Z(\alpha))](\bar{k}_\shortparallel \delta\bar{\Phi}-\bar{\omega}\delta\bar{A}_\shortparallel )$.  

The pressure term in Eq.~\eqref{eq:dAdt_slab1d} is necessary to yield the correct dispersion relation of the Alfv\`en wave. Without the pressure term, the dispersion in this $A\&\Phi$ model is,
\begin{eqnarray}
    D_{{\rm w/o }\; p}=1+\frac{2}{k_\perp^2\rho_N^2 \bar{T}} [1+\alpha Z(\alpha)] 
    \frac{k_\perp^2 d_{\rm e}^2}{k_\perp^2 d_{\rm e}^2+1}\;\;,
\end{eqnarray}
which does not recover the Alfv\`en wave. Namely, although Eq.~\eqref{eq:dAdt_slab_nopressure} can be used for solving $\delta A_\shortparallel ^{\rm s}$, it can not be used as the $\delta A_\shortparallel $ equation without the additional $\delta A_\shortparallel ^{\rm h}$ equation to recover the Alfv\`en wave dispersion relation. 
On the contrary, Eq.~\eqref{eq:dAdt_slab1d} already gives a physical solution of the Alfv\`en wave even without solving the additional $\delta A_\shortparallel ^{\rm h}$ equation; while the $\delta A_\shortparallel ^{\rm h}$ equation can take into account other terms such as the non-uniform term ($\delta{\bm v}\cdot\nabla f_0$). 

\section{Numerical results and analyses}
\label{sec:results}
The 1D model has been implemented using the particle-in-Fourier method. To initialize the marker coordinates and parallel velocities $(x, v_\shortparallel )$, 
we employ the Hammersley sequence to generate two-dimensional quasi-random numbers. 
This choice ensures reproducibility while reducing errors associated with marker noise. Uniform distribution is adopted in the $x$ direction.  The distribution along $v_\shortparallel $ is either uniform or Maxwellian with a specific transformation from the Hammersley sequence, depending on the choice of the simulation. The marker weight is calculated according to $w_{0,p}=g(x_p,v_{\shortparallel ,p})/f(x_p,v_{\shortparallel ,p})$ for each marker ``$p$'', where $g$ and $f$ denote the distribution functions of the markers and physical particles, respectively. Compared with purely random sampling, the use of the Hammersley sequence yields a discrepancy that scales as $\propto 1/N_p$, rather than $\propto 1/\sqrt{N_p}$, where $N_p$ is the marker number, thereby ensuring faster convergence toward a uniform continuum of markers \cite{niederreiter1992random}.

\subsection{Electrostatic and electromagnetic limits of the dispersion relation}
\label{subsec:es_em}
\begin{figure}[t]
\centering
\includegraphics[width=0.4\textwidth]{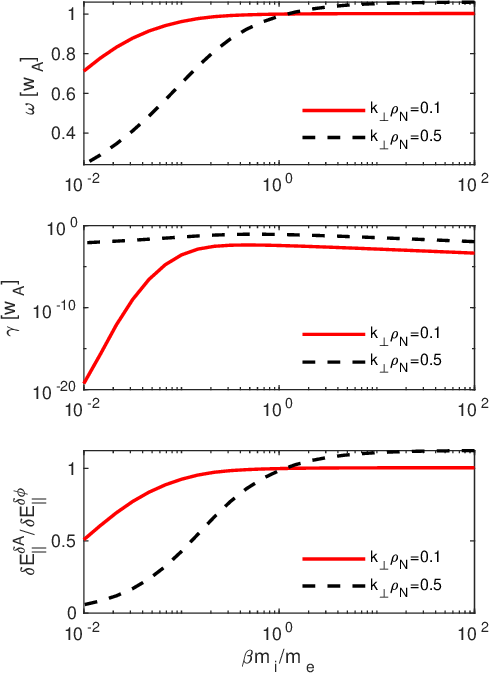}
\caption{The frequency (upper), the damping rate (middle), and the polarization (lower) of the shear Alfv\`en wave versus $\beta/\bar{m}_{\rm e}$ (Section \ref{subsec:es_em}). }
\label{fig:polarization}
\end{figure}

The 1D model in Eq.~\eqref{eq:dispersion_slab0} captures various physics properties for electrostatic and electromagnetic studies. As shown in Fig.~\ref{fig:polarization}, for $\beta m_{\rm i}/m_{\rm e}>10$, the wave frequency (top frame) is close to the Alfv\`en frequency $\omega_{\rm A}$, while as $\beta m_{\rm i}/m_{\rm e}$ decreases, the frequency becomes lower than $\omega_{\rm A}$. As $k_\perp\rho_N$ increases, the contribution from the ion polarization increases, leading to a larger deviation of $\omega$ from the Alfv\`en frequency $\omega_{\rm A}$ for small $\beta m_{\rm i}/m_{\rm e}$. The damping rate (middle frame) increases first and decreases as $\beta m_{\rm i}/m_{\rm e}$ increases. Larger $k_\perp\rho_N$ leads to a stronger damping.
The polarization is calculated as follows,
\begin{eqnarray}
\label{eq:polarization}
    \frac{\delta{E}_\shortparallel ^{\delta A}}{\delta{E}_\shortparallel ^{\delta \phi}} 
    \equiv
    \frac{|\partial_t\delta A|}{|{\bf b}\cdot\nabla\delta \phi|} 
    = \frac{\overline{\omega}^2}{\overline{\omega}_A^2} \;\;,
\end{eqnarray}
where $\overline{\omega}_A^2=\overline{k}_\shortparallel ^2/\beta^2$.
The polarization (lower frame), according to Eq.~\eqref{eq:polarization}, is electrostatic in the low $\beta m_{\rm i}/m_{\rm e}$ limit, which corresponds to the ``$\omega_{\rm H}$'' mode whose frequency is $\omega_{\rm H}=(k_\shortparallel /k_\perp)\sqrt{m_{\rm i}/m_{\rm e}}\omega_{{\rm  c}i}=\omega_{\rm A}\sqrt{\beta m_{\rm i}/ m_{\rm e}}/(k_\perp\rho_{\rm i})$ \cite{lee1987gyrokinetic}. In the following particle simulations, we choose cases with high and low $\beta m_{\rm i}/m_{\rm e}$ values to test the performance in  the electrostatic and electromagnetic limits, respectively. 

\subsection{Comparison of particle simulation and theory}
\label{subsec:compare_theory}
\begin{figure}[t]
\centering
\includegraphics[width=0.48\textwidth]{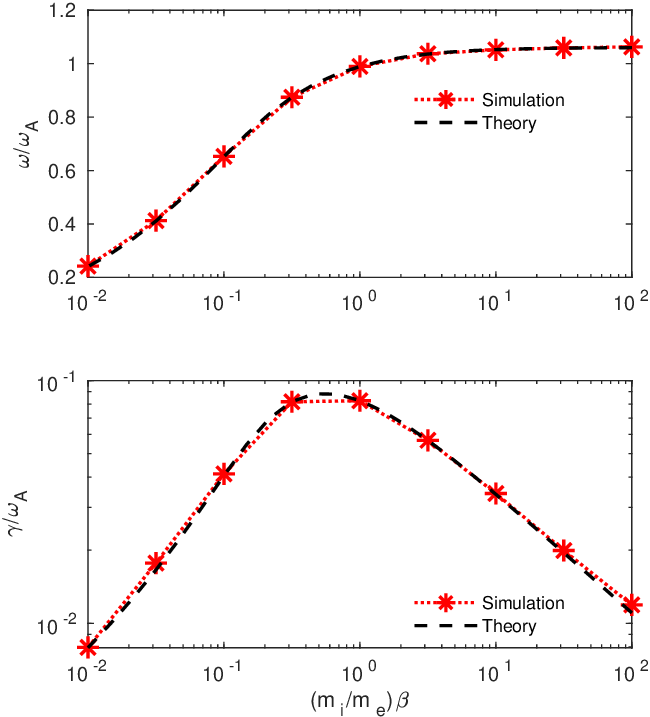}
\caption{The frequency (upper) and the damping rate (lower) of the shear Alfv\`en wave versus $\beta/\bar{m}_{\rm e}$. The parameters are $\bar{m}_{\rm e}=0.001$, $\rho_N=0.01$, $k_\perp=50$ (Section \ref{subsec:compare_theory}). }
\label{fig:compare_theory}
\end{figure}

The pure $v_\shortparallel $ form is implemented using Eqs.~\eqref{eq:qn_slab}, \eqref{eq:ampere_slab}, and \eqref{eq:dfdt_slab}. 
The markers are pushed along perturbed trajectories but the nonlinear correction is negligible since the initial perturbation is small ($\delta f/f_0\sim 10^{-8}$). 
The simulation domain is $x\in[0,1]$ m, with $k_\shortparallel =2\pi$. The parameters are $\rho_N=0.01$ m, $k_\perp=50/{\rm m}$, $m_{\rm e}/m_{\rm i}=0.001$. The marker number is $N_p=10^5$. The time-step size is $\Delta t=T_{\rm A}/64$, except for the last two points at high $\beta$, $\Delta t=T_{\rm A}/128$, $T_{\rm A}/256$, where $T_{\rm A}=2\pi/\omega_{\rm A}$, and $\omega_{\rm A}$ is the Alfv\`en frequency. For the base case, $\beta=0.001$, $1/(k_\perp^2d_{\rm e}^2)=4000\beta=4$. The results using the pure $v_\shortparallel $ model and the comparison with the theoretical result are demonstrated in Fig.~\ref{fig:compare_theory}, where the highest $(\beta m_{\rm i}/m_{\rm e})_{\rm max}=100$, corresponding to $1/(k_\perp^2d_{\rm e}^2)=25000\beta=4000$. A good agreement between the simulation  and the theory is achieved even for the weakly damped case with $\gamma/\omega\sim 1\%$.

\subsection{Features of various schemes}
\label{subsec:scheme_features}
\begin{figure*}
\centering
\includegraphics[width=0.3\textwidth]{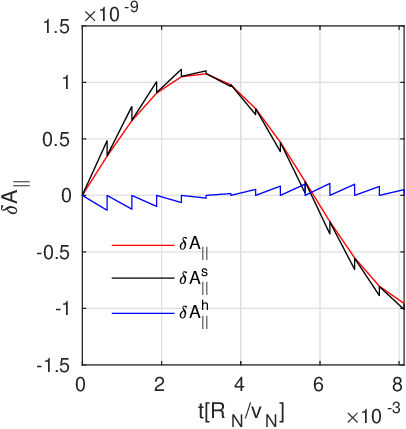}
\includegraphics[width=0.3\textwidth]{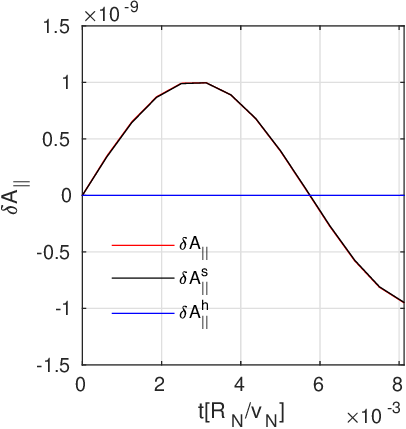}
\includegraphics[width=0.3\textwidth]{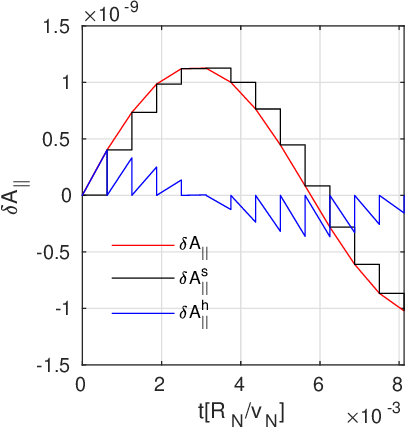}
\includegraphics[width=0.3\textwidth]{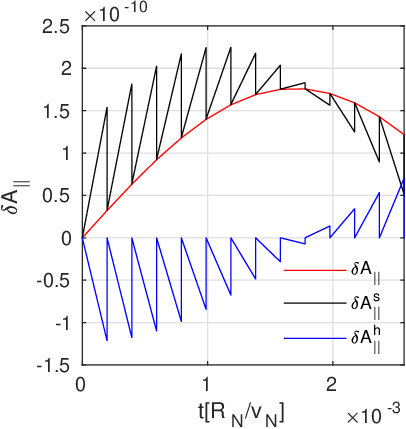}
\includegraphics[width=0.3\textwidth]{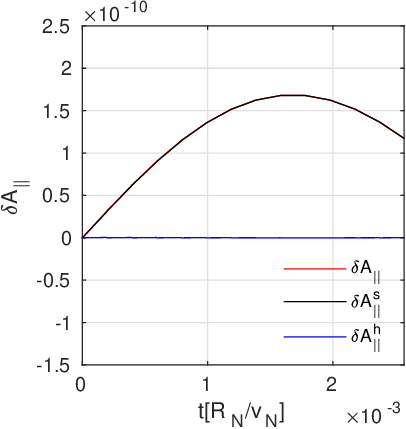}
\includegraphics[width=0.3\textwidth]{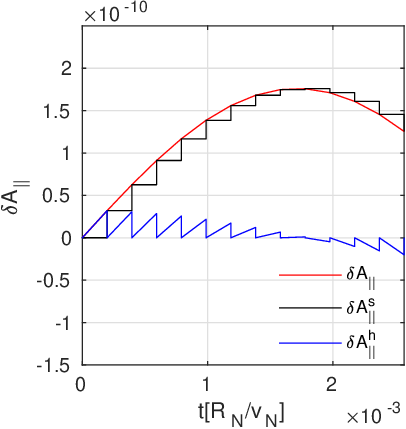}
\caption{The time eovlution of $\delta A_\shortparallel ^{\rm s}$, $\delta A_\shortparallel ^{\rm h}$, and $\delta A_\shortparallel $ for the mixed variable-pullback scheme with ideal Ohm's law (left, scheme 1), with non-ideal Ohm's law (middle, scheme 2a) and without Ohm's law (right, scheme 3a). Upper frame: $\beta=0.001$. Lower frame: $\beta=0.0001$. (Section \ref{subsec:scheme_features}). }
\label{fig:evolution_A}
\end{figure*}

We focus on the low $\beta$ case with $k_\perp=20$ and $\bar{m}_{\rm e}=0.001$, corresponding to a weakly electromagnetic effect. Other parameters are the same as those in Section \ref{subsec:compare_theory}. 
The time evolution of $\delta A_\shortparallel ^{\rm s}$, $\delta A_\shortparallel ^{\rm h}$, and $\delta A_\shortparallel $ is analyzed in Fig.~\ref{fig:evolution_A}. The time-step size $\Delta t=T_{\rm A}/16$, and the initial perturbation $w_p=A_w\sin(k_\shortparallel x)$ where $A_w=10^{-8}$. Two sets of cases are analyzed with $\beta=0.001$ (upper frame) or $\beta=0.0001$ (lower frame). The discontinuity of the curves corresponds to the pullback operation. For $\beta=0.001$, using the mixed variable-pullback scheme with the ideal Ohm's law, the local time derivative of $\delta A_\shortparallel ^{\rm s}$ is already far away from that of the physical solution in $\delta A_\shortparallel $. However, the overall time evolution of $\delta A_\shortparallel ^{\rm s}$ follows that of the physical solution $\delta A_\shortparallel $ due to the pullback operation. For $\beta=0.0001$, the local time derivative of $\delta A_\shortparallel ^{\rm s}$ is too far away from that of the physical solution in $\delta A_\shortparallel $. As a result, even with a pullback, $\delta A_\shortparallel ^{\rm h}$ is significant. 

For the mixed variable-pullback scheme with the non-ideal Ohm's law, the time evolution of $\delta A_\shortparallel ^{\rm s}$ is almost identical to that of the physical solution $\delta A_\shortparallel $, and $\delta A_\shortparallel ^{\rm s}$  is from the small corrections due to the high-order nonlinear and nonuniform equilibrium terms as well as the numerical errors. 
The time evolution of schemes 2 and 2a is very similar since $\delta A_\shortparallel ^{\rm h}$ is much smaller than $\delta A_\shortparallel ^{\rm s}$. 

For the mixed variable-pullback scheme without Ohm's law, the time derivative of $\delta A_\shortparallel ^{\rm s}$ is zero since Ohm's law is not solved. $\delta A_\shortparallel ^{\rm h}$ is pulled back to zero at the end of each step, which is expected to reduce the particle noise. Without the pullback, the scheme is reduced to the traditional pure $p_\shortparallel $ scheme (Scheme 3b), for which $\delta A_\shortparallel ^{\rm s}=0$ and $\delta A_\shortparallel =\delta A_\shortparallel ^{\rm h}$. 

\subsection{Convergence analyses}
\label{subsec:convergence}
While the 1D model is simplified, it includes the necessary ingredients to test the performance of different schemes in the electrostatic limit or the electromagnetic limit. We use $\rho_N=0.01$ m, $k_\perp=50/{\rm m}$, $m_{\rm e}/m_{\rm i}=0.001$. Three cases are considered for the convergence analyses. 
\begin{enumerate}
    \item The high $\beta$ case that is close to the MHD limit, with $\beta m_{\rm i}/m_{\rm e}=100$ and $\delta E_\shortparallel ^{\delta A_\shortparallel }/\delta E_\shortparallel ^{\delta\phi}\approx 1.12$ (defined in Eq.~\eqref{eq:dAdt_slab1d}). Note that the finite $k_\perp\rho_N$ makes $\delta E_\shortparallel ^{\delta A_\shortparallel }/\delta E_\shortparallel ^{\delta\phi}$ away from $1$, as shown by the black dashed lines in Fig.~\ref{fig:polarization}. 
    \item The moderate $\beta$ case with $\beta m_{\rm i}/m_{\rm e}=1$ and $\delta E_\shortparallel ^{\delta A_\shortparallel }/\delta E_\shortparallel ^{\delta\phi}\approx 0.986$.
    \item The low $\beta$ case that corresponds to the electrostatic limit with $\beta m_{\rm i}/m_{\rm e}=0.01$ and $\delta E_\shortparallel ^{\delta A_\shortparallel }/\delta E_\shortparallel ^{\delta\phi}\approx 0.0579$.
\end{enumerate}

\begin{figure*}
\centering
\includegraphics[width=0.48\textwidth]{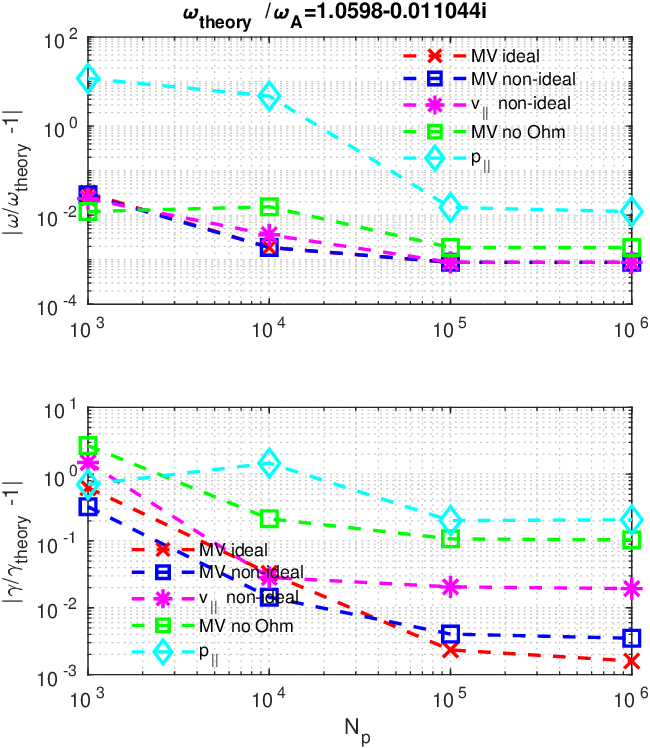}
\includegraphics[width=0.48\textwidth]{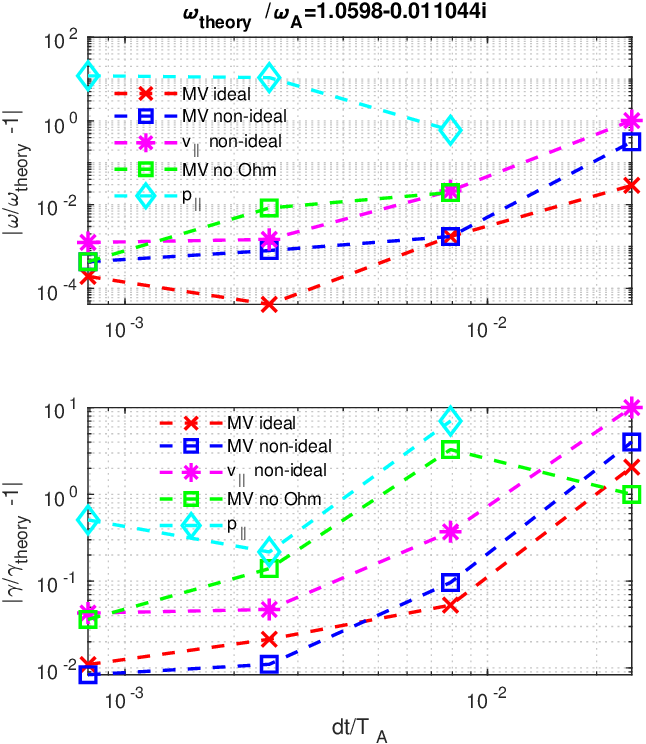}
\caption{The frequency (upper) and the damping rate (lower) of the shear Alfv\`en wave versus the marker number with $\Delta t/T_{\rm A}=1/256$ (left) and versus the time-step size with $N_p=10^4$ (right) for different schemes in the MHD limit ($\beta/\bar{m}_{\rm e}=100$) (Section \ref{subsec:convergence}). }
\label{fig:converge_em}
\end{figure*}

\begin{figure*}
\centering
\includegraphics[width=0.48\textwidth]{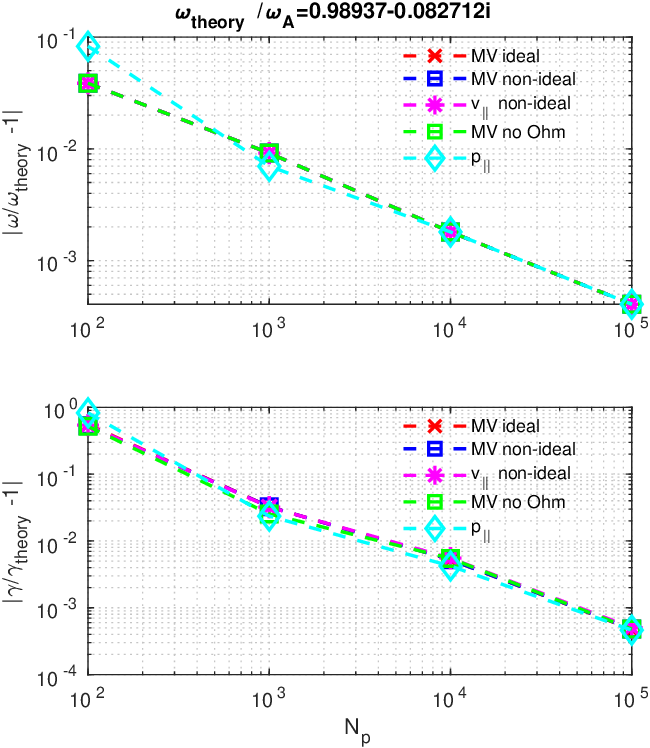}
\includegraphics[width=0.48\textwidth]{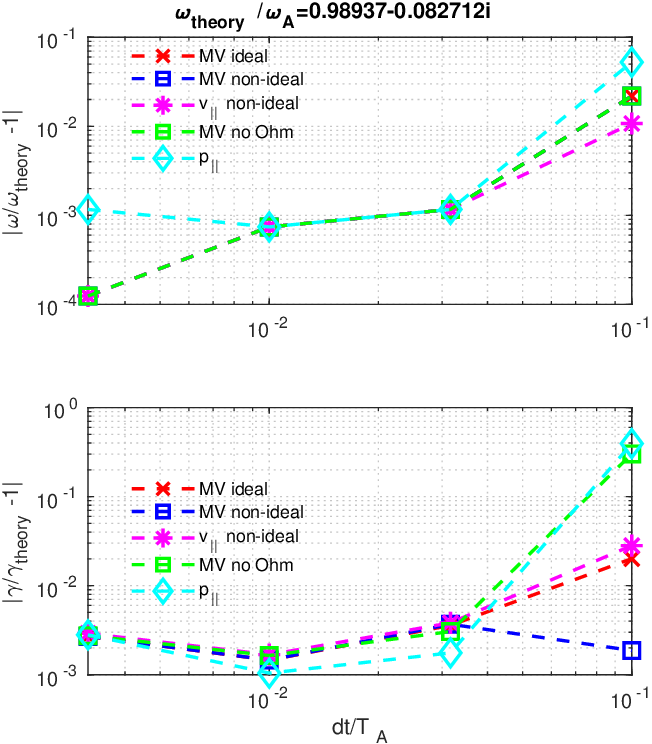}
\caption{The frequency (upper) and the damping rate (lower) of the shear Alfv\`en wave versus  marker number with $\Delta t/T_{\rm A}=1/64$ (left) and versus the time-step size with $N_p=10^4$ (right)  for different schemes for moderate $\beta$ ($\beta/\bar{m}_{\rm e}=1$) (Section \ref{subsec:convergence}). }
\label{fig:converge_sm}
\end{figure*}

\begin{figure*}
\centering
\includegraphics[width=0.48\textwidth]{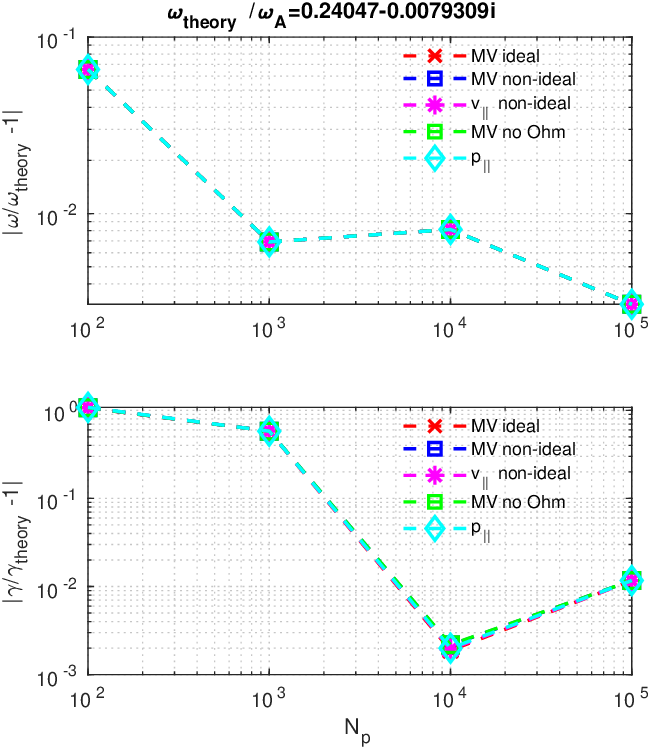}
\includegraphics[width=0.48\textwidth]{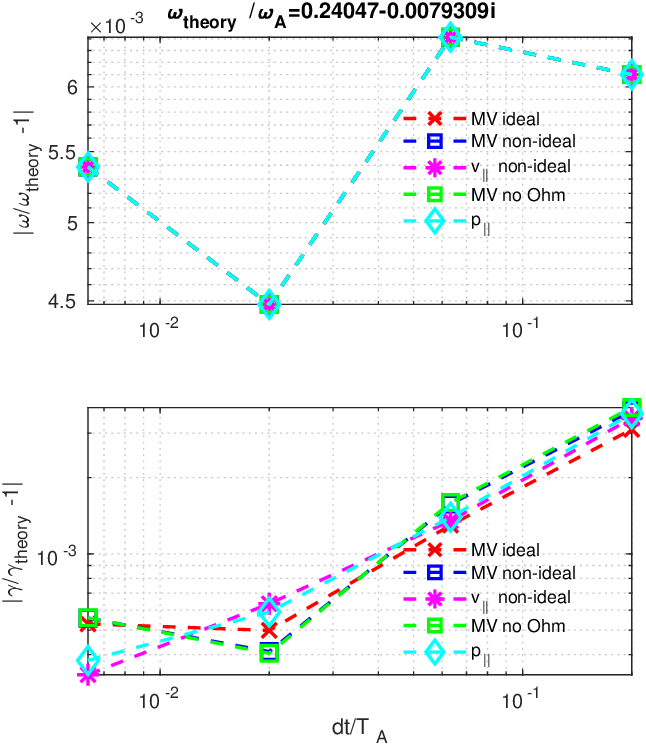}
\caption{The frequency (upper) and the damping rate (lower) of the shear Alfv\`en wave versus marker number with $\Delta t/T_{\rm A}=1/16$ (left) and versus the time-step size with $N_p=10^4$ (right)  for different schemes in the electrostatic case ($\beta/\bar{m}_{\rm e}=0.01$) (Section \ref{subsec:convergence}). }
\label{fig:converge_es}
\end{figure*}

The growth rate and the frequency are calculated for different marker numbers and time-step sizes. The relative errors of the frequency and the damping rate, defined as $\varepsilon_\omega=|\omega/\omega_{\rm theory}-1|$ and $\varepsilon_\gamma=|\gamma/\gamma_{\rm theory}-1|$, are shown in Figs.~\ref{fig:converge_em} to ~\ref{fig:converge_es}. 

In the high $\beta$ case as shown in Fig.~\ref{fig:converge_em}, the left panel shows the convergence with respect to the number of markers $N_p$. The convergence of the pullback scheme with ideal Ohm's law or non-ideal Ohm's law is similar. In addition, with the non-ideal Ohm's law, the pure $v_\shortparallel $ and the mixed variable scheme with no Ohm's law also show a good convergence, but the damping rate is significantly less accurate than the former two. The pure $p_\shortparallel $ scheme gives much worse convergence than the other four schemes. While the pure $p_\shortparallel $ scheme suffers from the cancellation problem significantly, the mixed variable-pullback scheme without Ohm's law is much better with only a pullback operation than the pure $p_\shortparallel $ scheme. 
The right panel of Fig.~\ref{fig:converge_em} shows the convergence of growth rate and the frequency for different time-step sizes $\Delta t$. 
The mixed variable-pullback scheme with ideal or non-ideal Ohm's law and the pure $v_\shortparallel $ scheme show similar convergence as $\Delta t$ changes. The mixed variable scheme without Ohm's law and the pure $p_\shortparallel $ scheme have worse convergence as $\Delta t$ increases. 

The convergence of different schemes in the moderate $\beta$ case is shown in Fig.~\ref{fig:converge_sm}. The frequency and the damping rate start to converge with the relative error lower than $10^{-2}$ at $N_p=10^4$ for all schemes. All schemes have  minor differences in the convergence of the total marker number (left) and the time-step size (right). 

The convergence in the low $\beta$ case is shown in Fig.~\ref{fig:converge_es}. The results start to converge at $N_p=10^3$ for all schemes. All schemes have the same convergence of the total marker number (left) and similar convergence of the time-step size (right). In the electrostatic limit, although the mixed variable-pullback scheme with the ideal Ohm's law leads to an artificial $\delta A_\shortparallel ^{\rm s}$ and $\delta A_\shortparallel ^{\rm h}$ as we demonstrated in Fig. \ref{fig:evolution_A} and discussed in Sec. \ref{subsec:scheme_features}, the magnitudes of $|\delta A_\shortparallel ^{\rm s}/\delta\phi|$ and $|\delta A_\shortparallel ^{\rm h}/\delta\phi|$ remain small and the results are insensitive to the treatment of electromagnetic variables. 

\section{Summary and outlook}
\label{sec:summary}
This work generalizes the mixed variable-pullback scheme through an extension of the solution for the symplectic part. The non-ideal Ohm's law has been integrated with the mixed variable-pullback scheme to provide a more physical solution to the symplectic part of the system. The pure $v_\shortparallel $ scheme is recovered by the generalized mixed variable-pullback scheme using the non-ideal Ohm's law. The main conclusions in this work are as follows.
\begin{enumerate}
    \item The pure  $v_\shortparallel $ scheme with the non-ideal Ohm's law is capable of simulating the electromagnetic Alfv\`en wave in the MHD limit with a small electron skin depth $1/(k_\perp^2d_{\rm e}^2)=\beta/(\bar{m}_{\rm e}k_\perp^2\rho_{\rm i}^2)=4000$. 
    \item The pure $v_\shortparallel $ scheme and the mixed variable-pullback scheme with the non-ideal Ohm's law give similar convergence as the mixed variable-pullback scheme for the electromagnetic 1D shear Alfv\`en wave problem.
    \item The mixed variable-pullback scheme without Ohm's law improves the traditional pure $p_\shortparallel $ scheme significantly, requiring fewer markers to achieve the same level of accuracy.
    \item The convergence of the time-step size and the marker number in the mixed variable-pullback scheme is not sensitive to the choice between the ideal or non-ideal Ohm's law. Both schemes are equally effective with minor differences in specific parameters.
\end{enumerate}

This work addresses various treatments for enhancing the performance of electromagnetic gyrokinetic codes. The mixed variable-pullback scheme without Ohm's law provides a straightforward improvement over the traditional pure Hamiltonian scheme, allowing the same accuracy with a smaller marker number and minimal modifications to both the model and implementation. The pure $v_\shortparallel $ and mixed variable-pullback schemes with the non-ideal Ohm's law provide alternative options for upgrading an electrostatic gyrokinetic code to the electromagnetic version, in addition to the well-known mixed variable-pullback scheme with ideal Ohm's law.

\begin{acknowledgments}
The simulations in this work were run on the TOK cluster and the MPCDF Viper/Raven supercomputers.  
The Eurofusion projects TSVV-8, ACH/MPG and TSVV-10 are acknowledged. 
This work has been carried out within the framework of the EUROfusion Consortium, funded by the European Union via the Euratom Research and Training Programme (Grant Agreement No 101052200—EUROfusion). Views and opinions expressed are however those of the author(s) only and do not necessarily reflect those of the European Union or the European Commission. Neither the European Union nor the European Commission can be held responsible for them.
\end{acknowledgments}

\end{document}